\documentclass{iopart}
\usepackage{epsfig}
\usepackage{cite}
\begin{document}
\jl{3}
\letter{Intrinsic hole localization mechanism in magnetic semiconductors}

\author{H. Raebiger,  A. Ayuela\dag\ and  R. M. Nieminen}
\address{Laboratory of Physics, Helsinki University of Technology, 02015 HUT, Finland}
\address{\dag\ Donostia International Physics Center (DIPC), 20018 San Sebastian/Donostia, Spain}

\begin{abstract}
The interplay between clustering and  exchange coupling in
magnetic  semiconductors for  the  prototype
(Ga$_{1-x}$,Mn$_x$)As is investigated
considering manganese concentrations $x$ of 1/16 and 1/32, 
which are in the interesting experimental range.
For $x\sim 6 \%$, we study all
possible  arrangements  of two Mn atoms on the Ga sublattice within  a  large supercell
and find the clustering of  Mn atoms at nearest-neighbour
Ga sites
energetically  preferred. 
As  shown by analysis of spin density and projected density-of-states,   
this  minimum energy  configuration localizes further one hole 
and  reduces  the effective  charge  carrier  concentration. Also  the
exchange coupling constant increases
to a value corresponding to lower Mn concentrations
with decreasing inter Mn distance.
\end{abstract}

Including spin information into semiconductor electronics
has enormous potential for new applications
(see Ref. \cite{prinz} for a review of magnetoelectronics).
Within this field of research,  
the III-V diluted magnetic semiconductors (DMS)
have opened up a new chapter,
in particular (Ga,Mn)As is a prominent candidate for a spintronics material.
This compound, grown by means of low-temperature molecular beam
epitaxy (LT-MBE), \cite{munekata,ohno-sc,ohno-ssc,ohno-mmm}
shows ferromagnetism \cite{munekata,ohno-sc,ohno-ssc,ohno-mmm,ohldag,hayashi,potashnik}
with a Curie temperature
as large as 110 K in the Mn concentration  range between $5\%$ and $10\%$ \cite{ohno-sc,ohno-ssc,ohno-mmm}.
The  ferromagnetism (FM) in (Ga,Mn)As is mediated by holes that are
antiferromagnetically (AFM) coupled to the Mn. 
In spite of the tremendous theoretical effort to explain the interplay of
magnetism and semiconductor properties
\cite{shirai01,shirai,ogawa,akai,ordejon,inoue,schilfgaarde,sanvito-apl,sanvito,Mac...},
a number of fundamental questions still
\begin{figure}[tb]
\begin{center}
\mbox{
\epsfig{figure=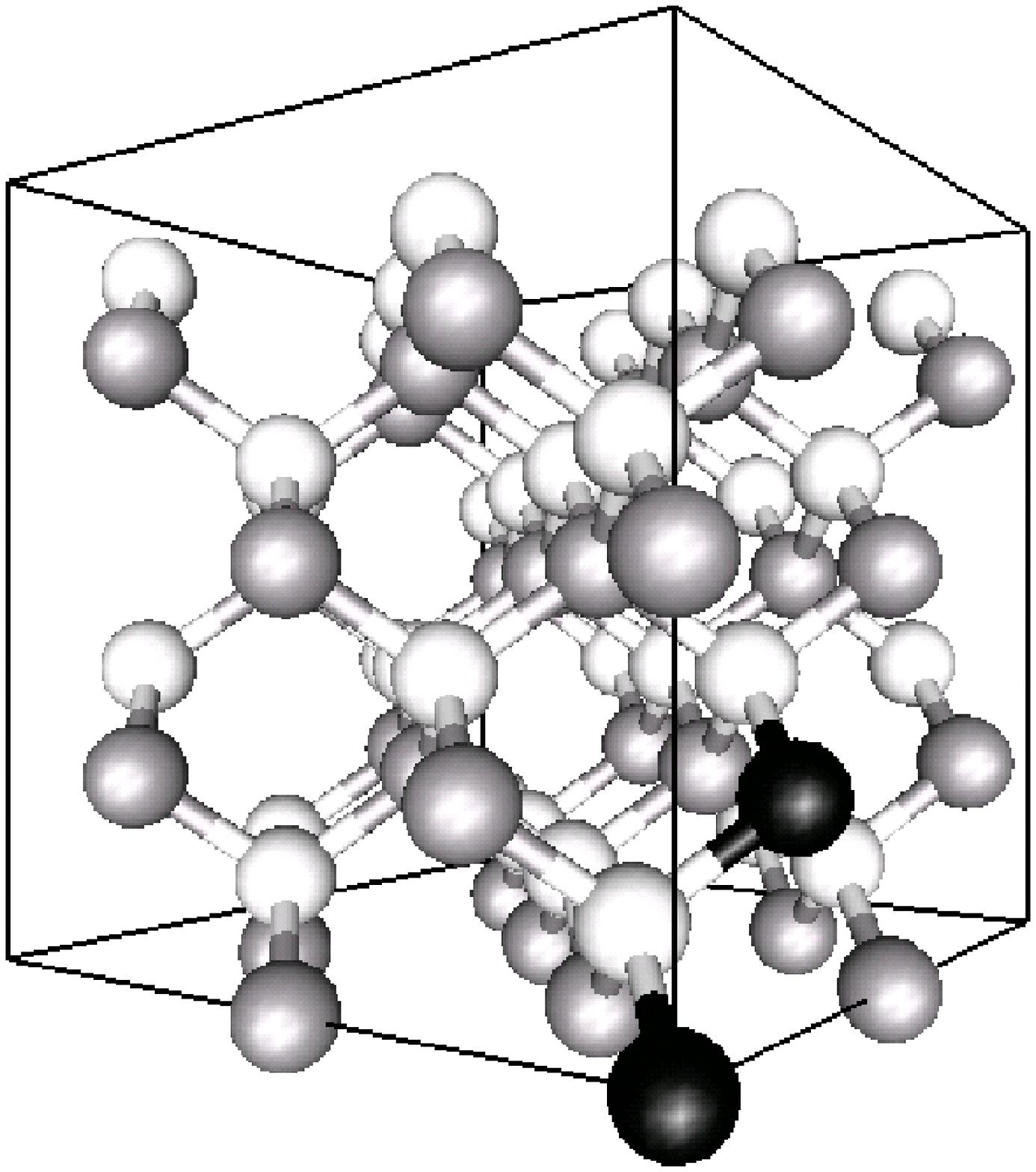,width=0.3\linewidth}
\epsfig{figure=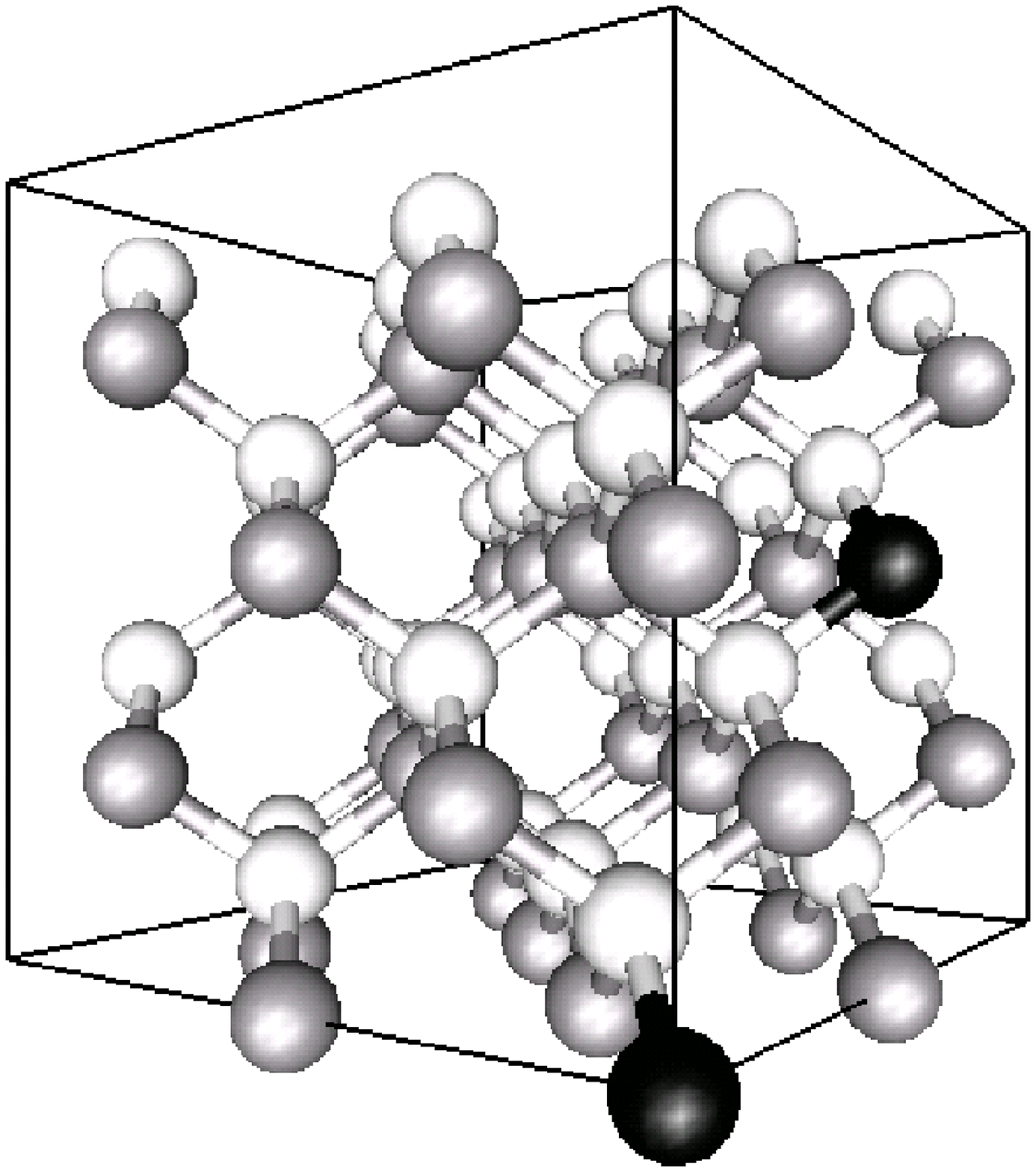,width=0.3\linewidth}
\epsfig{figure=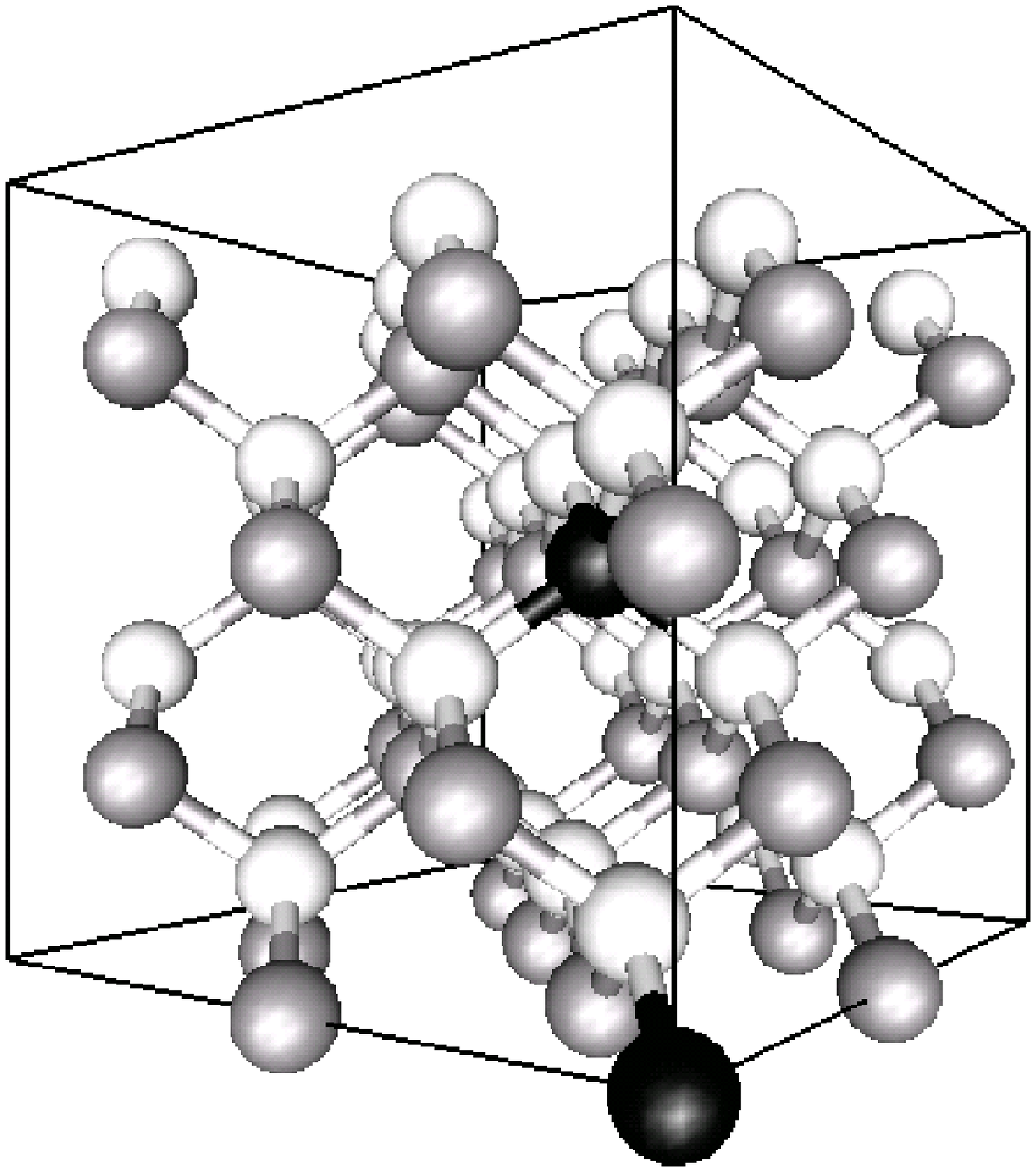,width=0.3\linewidth}
}\\
\small{[n-n, ground state (211 meV)] [face, +42 meV (217 meV)] [n-h, +100 meV (160 meV)]}\\
\mbox{
\epsfig{figure=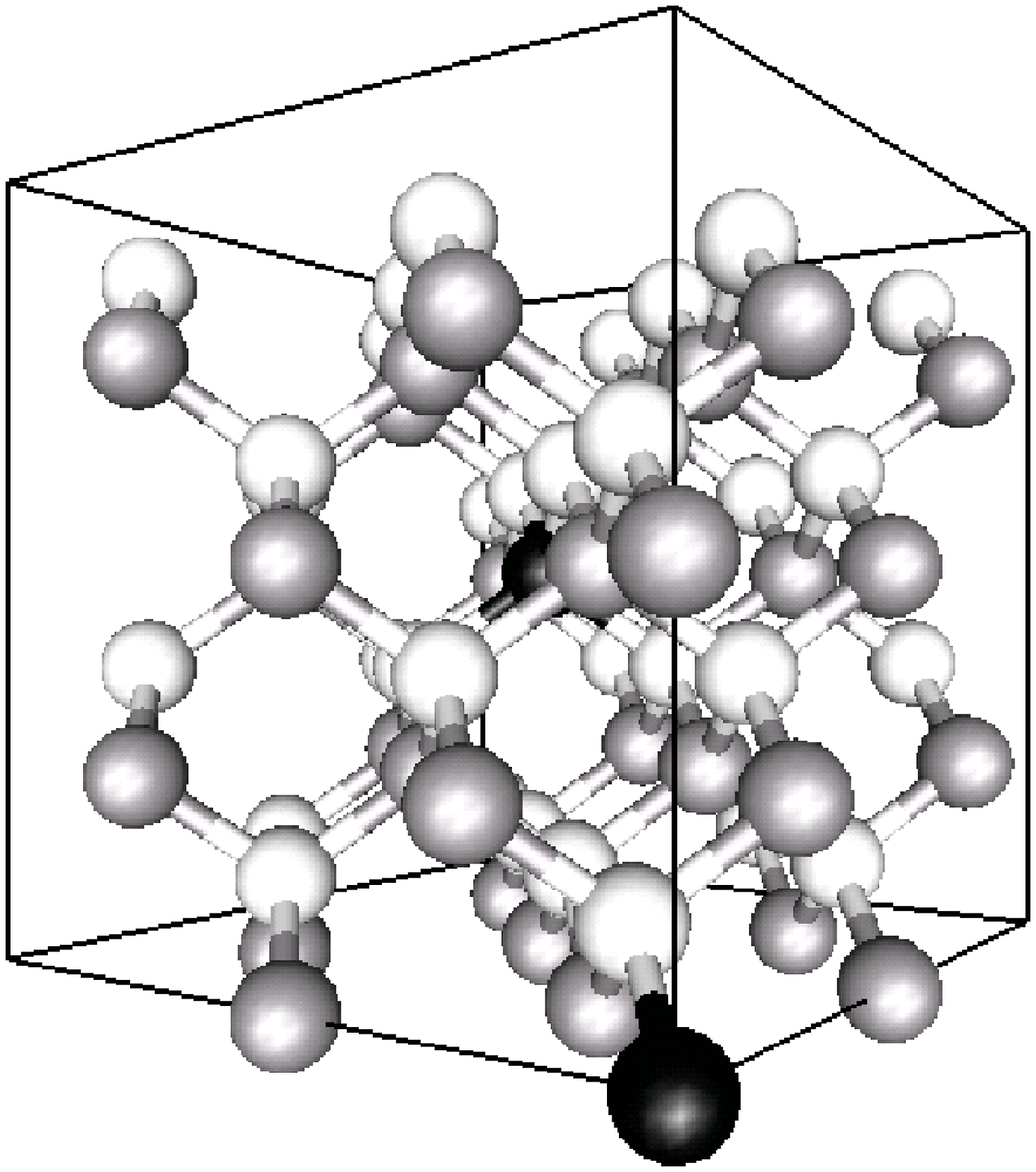,width=0.3\linewidth}
\epsfig{figure=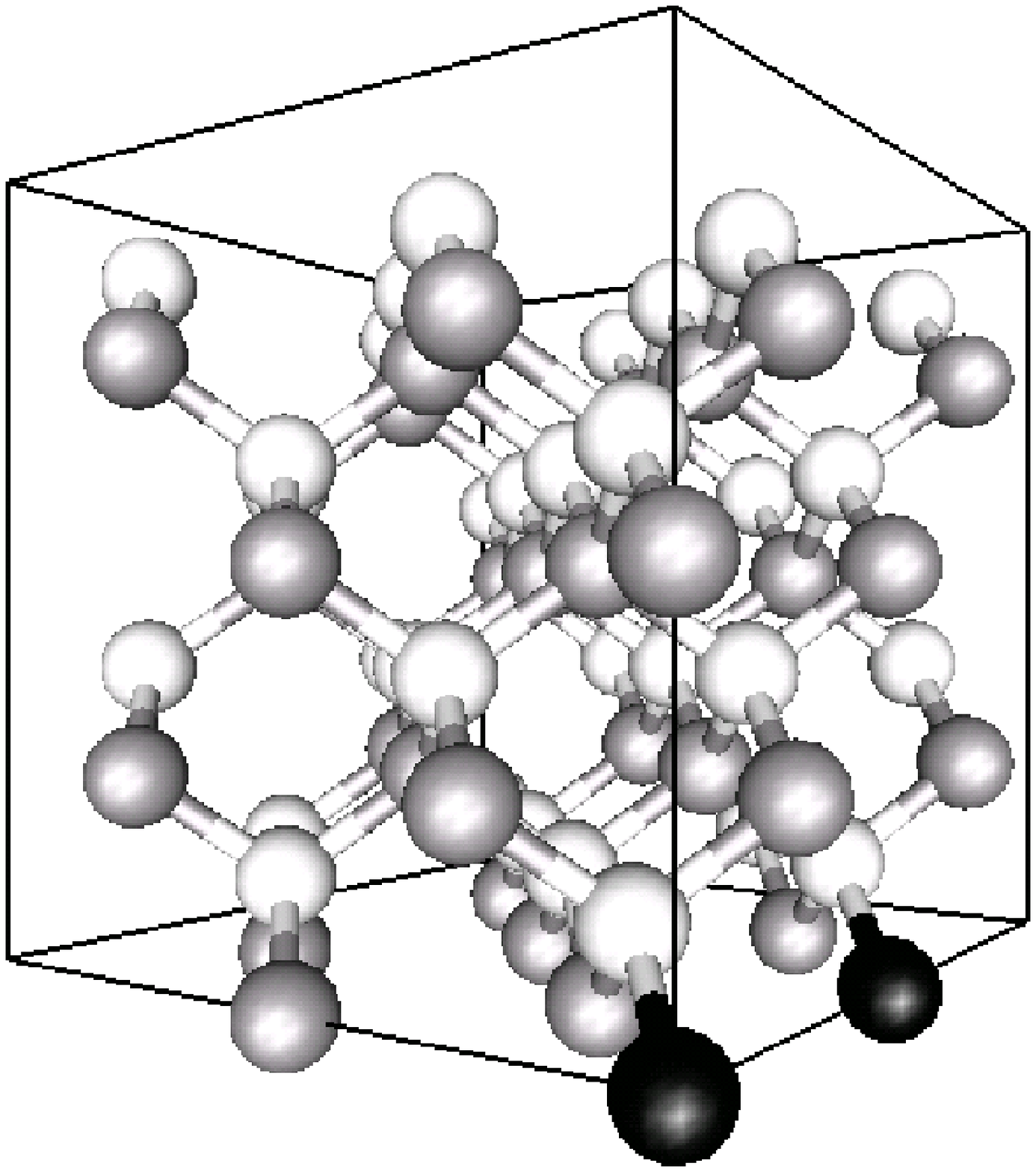,width=0.3\linewidth} }\\ 
\small{[homog., +114 meV (171 meV)] [edge, +184 meV (67 meV)]}
\end{center}
\caption{Structures of the (Ga$_{1-x}$,Mn$_x$)As system with x $\sim$ 6 \% and
successive stages of clustering. Ga atoms are shown as grey, Mn atoms as black,
and As as white. They are ordered according to the energy in meV respect to the
ground state, given by the number below the structures. In parenthesis
 the antiferromagnetic-ferromagnetic AFM-FM energy differences 
$\Delta_{AFM-FM}$
are also provided.}
\label{fig:geometry}
\end{figure}
remain unresolved. 
Such is the so-called hole compensation problem:
although the substitutional Mn is a single acceptor and should introduce one hole,
the observed hole concentrations are at least an order of magnitude smaller than
the Mn concentration \cite{experiments}. 
This hole compenation has been abscribed to 
intrinsic defects such as Mn interstitials \cite{erwin}
and As antisites \cite{akai}.
However, the antisite concentrations are small, interstitial Mn 
has a high formation energy
compared with substitutional Mn \cite{erwin,mahadevan},
and the interstitial Mn has a rather low migration barrier
so annealing will reduce interstitial concentrations
\cite{edmonds}.
Still,  after growth condition optimization and thermal annealing
the hole compensation cannot completely be eliminated,
and long-time annealing  reduces further the hole concentration
\cite{potashnik}.
We focus on the hypothetical defect-free (Ga$_{1-x}$,Mn$_x$)As
and consider hole compensation due to 
intrinsic properties of the substitutional Mn and the spin-polarized holes, 
bonding and localization effects.

In this Letter we present first principles calculations for magnetic semiconductors.
Within the density-functional theory we test all-electron methods,
the full potential linearized augmented plane wave (FLAPW) method \cite{Wien97}
and the projector augmented-wave (PAW) method \cite{vasp,newPAW},
as well as a plane wave pseudopotential (PWPP) method \cite{vasp}.
We also test the generalized gradient approximation (GGA)
beyond the local spin density (LSD) approximation\cite{another,lsd}
for the exchange-correlation functional.
We then investigate Mn clustering in (Ga,Mn)As within the supercell approach 
(see Fig.\ref{fig:geometry} describing the considered cases), which gives
us hints as to comment on hole localization. 
Based on the clustering results, we reexamine  the findings of 
previous calculations that address the role of the exchange 
coupling parameter within mean-field theories.

As a benchmark we study zincblende MnAs which 
is a bulk equivalent to
the substitutional Mn in GaAs
\cite{shirai01,ordejon}.
To ensure convergence in energy and
magnetic moment with respect to k-points and cutoff parameters,
especially the structure of the density-of-states (DOS) was monitored. 
\begin{table}
\caption{Equilibrium lattice constant and 
electronic behaviour for the 
zincblende MnAs structure. The calculations have been done
within several density-functional methods with different treatments: i) for the
atom potential (FLAPW- and PAW- full potential; PWPP- pseudopotentials); ii)
or for
the correlation approximation (GGA- generalized gradient approximation; 
LSD- local spin density approximation).}
\label{tab:test}
\begin{tabular}{l|c|c|c|c}
 & \multicolumn{2}{c|}{Lattice constant(\AA)} & \multicolumn{2}{c}{equilibrium behaviour} \\
 & GGA & LSD & GGA & LSD \\
\hline
FLAPW           & 5.63  & 5.34  &metallic & metallic \\
        & - & 5.85 \cite{shirai01}   & - & half-metal \\
PWPP            & 5.72  & 5.39  & half-metal & metallic \\
        & - & 5.6-5.7 \cite{ordejon}   & - & metallic \\
\hline
{PAW}               & { 5.61 } & { 5.33 } & { metallic} & { metallic} \\
\hline
\end{tabular}\\
\end{table}
We see in Table \ref{tab:test} obvious differences between our 
full-potential and pseudopotential calculations for the lattice constants 
between   GGA and LSD 
results. They show errors
of the order of 0.1 \AA, which are far from negligible because one is near 
 the transition from metallic behaviour to  a half-metallic state.
Both the lattice constants as well as the electronic behaviour show differences
with respect to previous calculations \cite{shirai,sanvito}.
The differences concerning pseudopotential calculations might be
ascribed to the used pseudopotentials, while the difference with 
previous full-potential calculations sounds more puzzling.
To correctly observe the metallic or half-metallic behaviour,
a denser mesh beyond a 
typical semiconductor k-point sampling is required.
Apart from the intrinsic pseudopotential differences,
the k-point consideration remains also as the main
difference with respect to previous
pseudopotential calculations (see Fig 4 in Ref. \cite{sanvito}, where the 
magnetization shows spurious jumps with the distance).
The tests clearly show that calculations must be carried out within all-electron methods
and the GGA.
The PAW calculations are computationally nearly as efficient as the PWPP ones,
so the PAW GGA approach is employed henceforth.
We have made simulations of Ga$_{1-x}$Mn$_x$As with x= 1, $\frac{1}{16}$, and 
$\frac{1}{32}$ where the notation of $x$ means $\frac{Mn-atoms}{As-sites}$ in the 
supercell. Naturally the size of the supercell is twice the number of
As sites. 
We only consider substitutional Mn in the Ga sublattice,
and focus on the clustering process of 2 Mn atoms in the high Curie temperature region
of $x \sim 6 \%$.
Here we include 2 Mn atoms in the 64 atom supercell (\textit{i.e.} $x = \frac{2}{32}$)
and calculate all five possible Mn arrangements 
corresponding to different Mn-Mn distances, as shown in Fig \ref{fig:geometry}. 
Again we test for convergence with respect to the number of k-points
and cutoff parameters: for the 64 atom supercell a 4x4x4 mesh
including the $\Gamma$ point is sufficient to sample
the Brillouin zone, and plane waves up to the cutoff of 275 eV give
converged results.
The lattice atoms are relaxed
so that the forces in the system are lower than 0.02 eV/\AA.

First we investigate the clustering process of Mn on the Ga sublattice.
In Fig. \ref{fig:geometry} the geometries are ordered according to their energy.
The energetic order  depends on the Mn-Mn atom distance 
except for the "edge" configuration. 
It is interesting to remark that the nearest neighbour 
configuration has the lowest energy which indicates clustering,
as also
predicted using the atomic sphere approximation 
for the potential \cite{schilfgaarde}. 
This clustering could play  a role in the magnetic order. 
Thus, the energetic differences $\Delta E _{AFM-FM}$
between ferromagnetic and antiferromagnetic configurations, 
\textit{i.e.} two Mn with parallel or antiparallel magnetic moments,
are also given in parenthesis in 
Fig. \ref{fig:geometry}. 
Tests on larger supercells showed that $\Delta E _{AFM-FM}$ is independent
of supercell size.
The  magnetic ground state of all the  
configurations is
ferromagnetic with
a magnetic moment of 4 $\mu_B$ 
per Mn atom in the unit cell, as for the homogenous case in Ref \cite{ordejon}. 
The $\Delta E _{AFM-FM}$  values
increase when decreasing the Mn-Mn distance.
For the edge configuration, the $\Delta E _{AFM-FM}$ is considerably smaller than in the other 
configurations. This configuration constitutes 
an exception because the As atoms are in a more open structure, 
which enables direct AFM coupling between the two Mn atoms.
Considering the other cases,
even for the larger distances, in the homogenous distribution, 
the values $\Delta E _{AFM-FM}$ are
large compared with  similar materials \cite{biplab}.
When neglecting the barriers involved in the process, the energies required by
the Mn 
to jump between different Ga sites are smaller than the energies required to 
flip the spins. In other words, the growth under strong magnetic fields
should affect the final    
atomic configuration. 

\begin{figure}[tb]
\begin{center}
\epsfig{figure=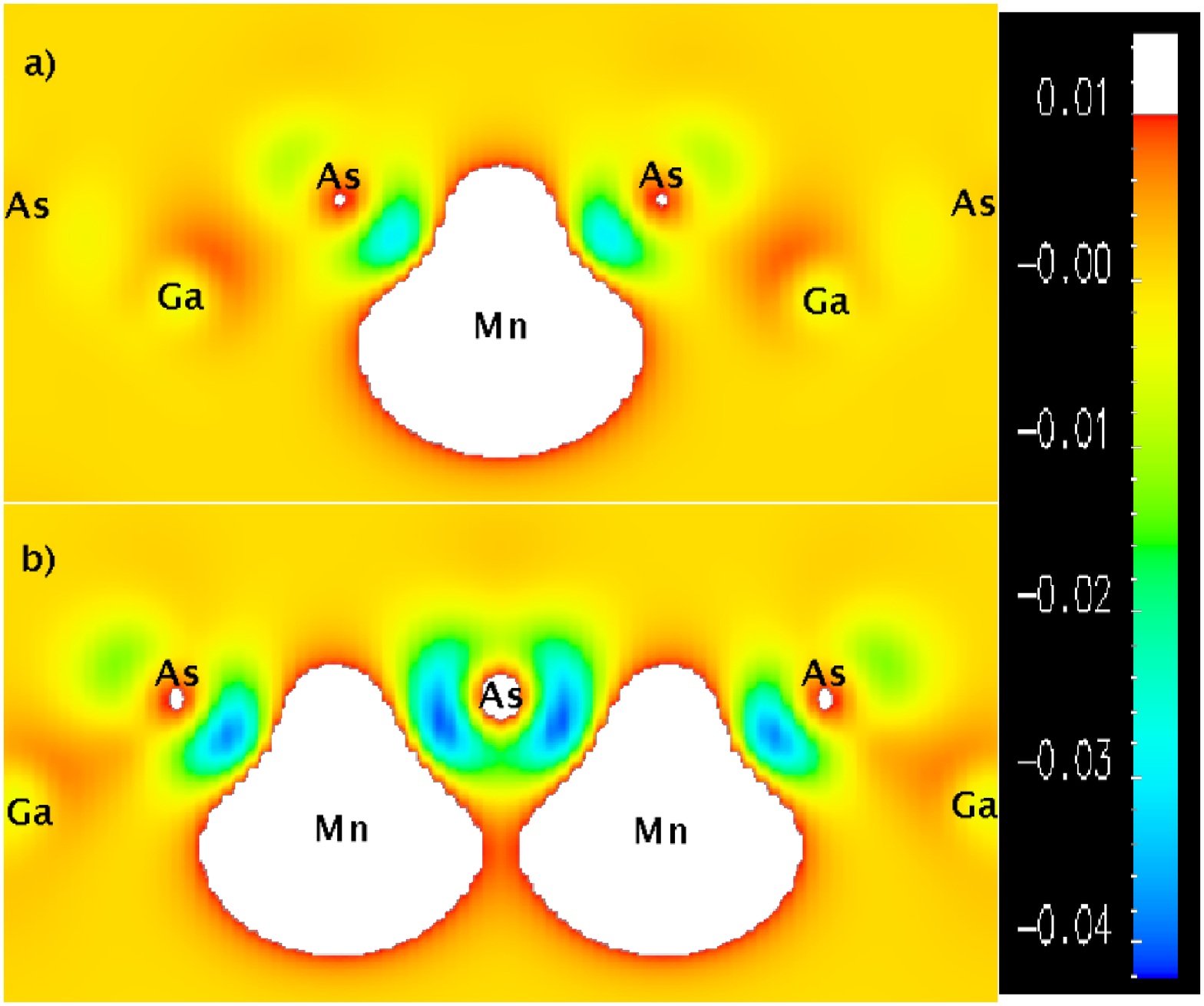,width=0.55\linewidth}
\epsfig{figure=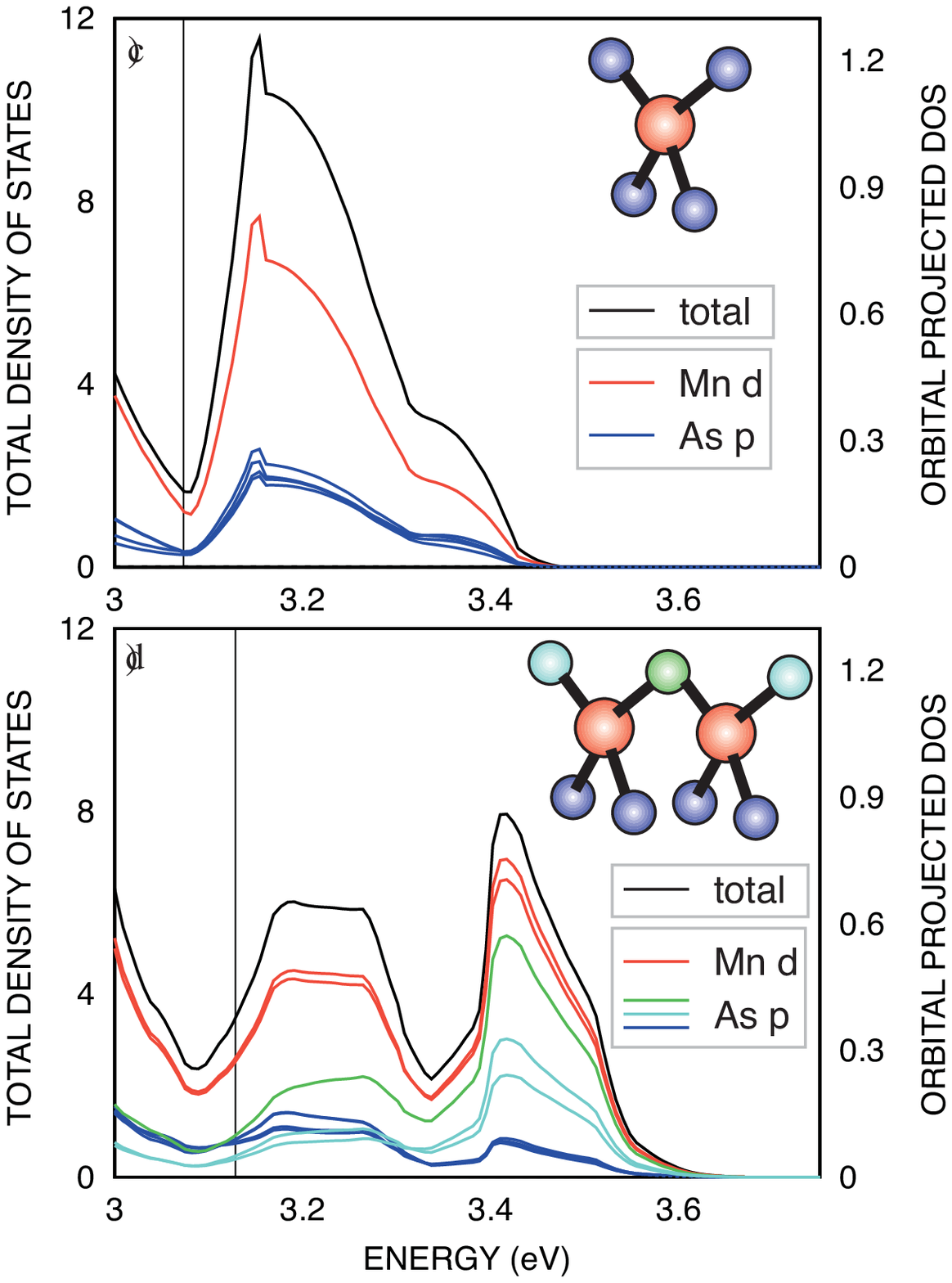,width=0.34\linewidth}
\caption{(Color) Spin density $\rho_\uparrow - \rho_\downarrow$ 
in the (110) plane around Mn (a and b) 
and majority spin density of states (c and d)
in the n-n and homogeneous case. Units in 
$\mu_B \rm{\AA}^{-3}$. Values over 0.1 $\mu_B \rm{\AA}^{-3}$ are truncated and given in 
white color. The DOS are given in states per supercell; black line for total DOS
and coloured lines for orbital projected DOS.
The red line represents Mn $d$ states, the green, turquoise and blue lines
represent the As $p$ states corresponding to different As sites,
as shown in inset.}
\label{fig:charge}
\end{center}
\end{figure}

Next we focus on the hole localization seen in the spin-density
as well as the density of states (DOS),
and look for quantifications by integrating over both spin-density and DOS.
The spin polarized charge density 
for the nearest neighbour 
and homogeneous case is presented in Fig. \ref{fig:charge} a and b. 
The negative 
spin density in the n-n configuration is more concentrated around the As nuclei than in the 
homogeneous one, especially around the As atom shared as first neighbour 
by both Mn atoms.
As FM is mediated by holes which are AFM coupled to Mn atoms,
the negative spin density points out the hole density.
Therefore this localization
of negative spin density
implies that the holes undergo strong localization in the
n-n case. 

This dimer-induced hole localization is further clarified
in the DOS shown in Fig. \ref{fig:charge} c and d,
where the $\rm{t_d}$-like hole states split into bonding and antibonding states.
Since we are discussing hole states, the behaviour is opposite to electronic states,
i.e. the bonding-like hole states are higher in energy 
(deeper in the gap, larger effective mass and higher localization).
From the orbital projected DOS (PDOS) we observe that the As $p$ states around the
in-between-As (green curve) 
provide a significant contribution to the DOS in the unoccupied region.
In homogeneous case the unoccupied states
correspond to antibonding $\rm{t^a}$ states that account for the Mn induced itinerant hole states
responsible for ferromagnetism (see Ref. \cite{sato}).
In the n-n configuration, the $\rm{t_d}$ symmetry break splits the $\rm{t^a}$ state
into an antibonding state and a bonding like state deeper in the gap.
The deeper state, according to PDOS analysis, is localized around the in-between-As,
and therefore we are left only with one itinerant $\rm{t^a}$ like hole state,
reducing the effective carrier concentration by one half.

For a quantative measure of how the holes are distributed on atomic orbitals
we integrate the PDOS over the unoccupied valence band states;
the absolute values depend on the basis set used, projection details etc.,
but relative values provide useful information when comparing different systems.
In the dimer configuration
we observe 0.1, 0.05 and 0.03 unoccupied $p$ states on the in-between-As (green),
the end-As (turquoise) and the vertical-As (blue), respectively following notation in Fig. 2.
In case of the homogeneous distribution,
we observe 0.04 unoccupied $p$ states on the As next to Mn.
It is interesting to compare the integrated hole densities with magnetization around As 
(\textit{i.e.} spin-density integrated inside sphere):
In the dimer configuration 
we observe values of -0.077 $\mu_B$, -0.032 $\mu_B$ and -0.009 $\mu_B$
around the in-between-As, end-As and vertical-As, respectively,
while in the homogeneous case the magnetization of the As next to Mn is -0.025 $\mu_B$.
The similar values for unoccupied As $p$ states and As magnetization
confirm that the negative spin-density actually is correlated with the hole density.
Notice than both values on the in-between-As get over twice the ones in the
homogenous case,which is a non-negligible quantity.
Thus this hole localization mechanism
seen both in spin density and DOS, 
combined with the lowered energy in the n-n
configuration is a significant addition to the previously known mechanisms
that reduce the effective carrier concentration. 
With this finding we give a possible explanation why only a fraction of the
Mn align ferromagnetically and why a large number of holes is lost, 
even when we are limited to only Mn in substitutional positions. 

\begin{figure}[tb]
\begin{center}
\epsfig{figure=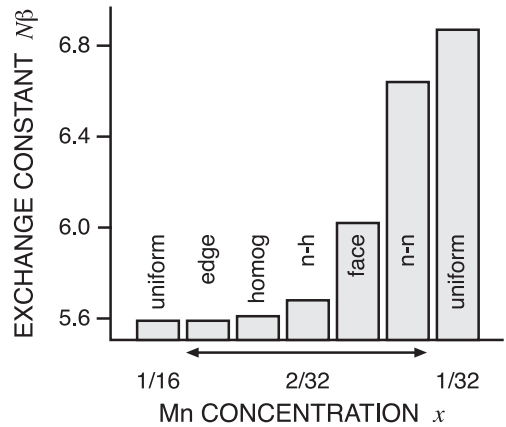,width=0.5\linewidth}
\caption{Exchange constant $N\beta$ (in eV) as a function of the Mn concentration x. Also
they are given as a function of the
geometrical arrangement for the Mn concentration x=2/32. 
}
\label{fig:splitting}
\end{center} 
\end{figure}

The change of DOS and the charge polarization for the different 
configurations drives us to address the 
same issues for the exchange coupling parameter $N\beta$, which is crucial 
in the calculation of thermodynamic properties \cite{Mac...}. 
Following Ref. \cite{ordejon}, the effect of the sp-d 
exchange on the band structure of the host semiconductor can be related
with the spin splitting
at the gamma point. Although
both conduction as well as valence band splitting can be considered, 
we concentrate only on the $N\beta$ parameter because it shows largest 
changes with the inclusion of  magnetism. 
The valence exchange coupling constant can be written as
\begin{displaymath}
N\beta=\frac{\Delta E^v} {x.\langle S \rangle} ,
\end{displaymath}
where $\Delta E^v$ is the valence band edge spin splitting at the $\Gamma$ point, and 
the mean field spin $\langle S \rangle$
is half of the computed magnetization per Mn ion.
Experimentally this constant is extracted from the exciton band that is
 spin split in optical magnetoabsorption experiments. 
Our estimated exchange couplings are given in Fig. \ref{fig:splitting}.

The parameter $N\beta $ for both concentrations shows values between 
5.6-6.8 eV.
With respect to  previous PWPP results, 5.48 eV for x=1/16 and 
7.34 eV for x=1/32 \cite{ordejon},
 $N\beta $ does not depend strongly 
on the Mn concentration. 
For the concentration $x=2/32$  the $N \beta$ parameter for several Mn configurations
increases as the  
longest distance between Mn atoms increases. 
Its value
approaches that of the smaller concentration
 x=1/32.
 It seems that the longest Mn-Mn distance
 sets the exchange
over  the whole semiconductor.
A simple mean-field model might after all be justified when re-interpreting the
exchange constant
in terms of holes ascribed both to Mn as well as Mn dimers,
\textit{i.e.} the Mn dimer should be treated like a single Mn impurity.
This is because the exchange is mediated by the itinerant holes,
and as seen in Fig. \ref{fig:charge} d, in the n-n case the two Mn 
as a total contribute only one
$\rm{t^a}$ like itinerant hole.
All these findings suggest that the breakdown of the mean field approximation
is not yet well established, \textit{i.e.} a mean field scheme may be justified by  including this clustering process. 

In conclusion, 
we see that disorder has strong effect on the ferromagnetic coupling. 
The preferred structural configuration 
is a dimer of two Mn occupying nearest neighbour Ga sites. 
This
configuration
localizes a hole and
explains in part the reduced carrier concentration
observed in experiments.
Further, our results suggest that the
analysis of the breakdown in the mean field
approximation  should be reconsidered.

\ack
This work has been supported by the Academy of Finland
(Centers of Excellence Program 2000-2005).  A. Ayuela is  supported by the EU TMR program (Contract
No. ERB4001GT954586). Computer facilities of the Center for
Scientific Computing (CSC) Finland are greatly acknowledged.
We thank K. Saarinen, K. Sato, H. Katayama-Yoshida, J. von Boehm,
M. J. Puska and M. Perez-Jigato 
for discussions during this work.

\end{document}